# Modeling Improved Fuzzy and Sliding-Mode for Single-Phase Inverter as Controllers in an EPS

Shou Qiao, Zhao Huio, Xio Tangue,
Huazhong University of Science and Technology

*Abstract*—In the last decades, the applications of power inverter increased rapidly. As a result, in spite of rectifier, an inverter with a high-power electronic oscillator has capability to convert direct current (DC) into alternating current (AC) in different forms. In this paper, a new fuzzy logic control (FLC) is suggested to an improved modeling inverter of a single-phase voltage source using LC filter and voltage sensor. Moreover, two sliding and fuzzy modes of non-linear voltage inverter controller are simulated and compared. The results proved the high efficiency and performance of the proposed method while reduce total harmonic distortion (THD) under linear loading conditions as well. All the data are applied on emergency power supply (EPS).

## I. Introduction

One of the most important device in electrical is inverter. Indeed, an inverter is a DC to AC device which is designed to producing AC power at a desirable current, frequency or voltage. In some applications such batteries, solar panels or fuel cells, the DC voltage is low. As a result, the inverters are used to convert DC into AC in devices with the ability to turn off AC power. One of the well-known illustration is conversion of electrical energy from a car battery to run a lap top, TV or cell phone. Most of the inverters operate in two different ways:

1. Converting DC input into AC output for reaching the main voltage via transformer and
2. Converting low voltage DC into high voltage DC and then DC wave into AC wave through Paul's H driver method [1-12].

In this paper, the relationships between the voltage and current supply of the inverter circuit outputs are taken into account. In a voltage source inverter (VSI), DC input voltage remains steadily constant while it is independent of the consumption load current increasing. An inverter is try to determine load voltage in case the load specifies the increased current. However, in a current source inverter (CSI), the output current is a constant value with a DC current input and load impedance determines the output voltage. Inverters are in a variety types with different outputs as follows:

Square wave, modified sinusoidal wave, pure/true sinusoidal wave (see fig.1) and square output wave with high harmonics [13].

Although, AC loads like motors and transformers are not efficient. The unit square wave is known as a pioneer development of inverter. A modified square wave or modified sinusoidal waveform is an inverter similar to an output square waveform, however, different in the output and gets close to zero volt for a period of time prior to negative or positive change. In fact, it is characterized by simplicity, inexpensiveness and compatibility with most electronic appliances. A pure sine wave inverter approximately generates an output sinusoidal waveform (harmonics less than 5%).

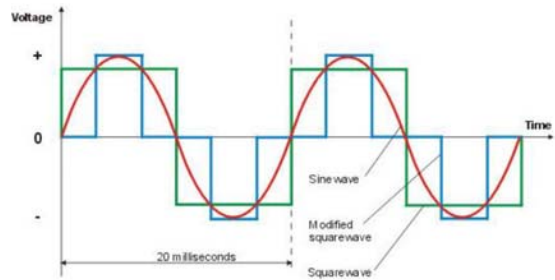

**Fig.1.** square output of the inverter

Most inverters have classical Proportional-Integral (PI) controllers, while are not suitable in many applications due to their output signals are not in pure sinusoidal waveform. This is because of harmonic distortion wave in/at loads which lead to a high power loss.

In this paper, a new Fuzzy logic controllers (FLCs) are applied to the inverter instead of the conventional PI controller. They results proved that it can proposed in an uncertain inputs while is more robust and non-linear manner. In fact, the commonly applied power inverters have a high THD, thereby being low in quality. Therefore, this paper tried to evaluate an inverter yield at an output waveform by using fuzzy logic control, which is a new method applied to controllers to minimize the output distortion so as to keep it at an acceptable level. The designed inverter output should be in pure sinusoidal form.

## II. CONVERTER MODELING

### 2.1. Depiction of a single-phase PWM Converter associated with a LC filter

The single-phase PWM converter for a load (fig.1) consists of two tools with bi-directional switches (insulated-gate bipolar transistor (IGBT) or metal–oxide–semiconductor field-effect transistor (MOSFET) with an anti-parallel diode) and plays a complementary role. The µ control signal, which is produced by a PWM generator, obtains values in a finite set of {1-, 1} *and summarizes µ1 and µ2 binary* orders from the two switching technique in µ = µ1 - µ2.

$$\begin{cases} 1: \mu1 = 1 \text{ and } \mu2 = 0 \rightarrow (K1, K'2) ON \text{ and } (K2, K'1) OFF \\ -1: \mu1 = 0 \text{ and } \mu2 = 1 \rightarrow (K2, K'1) ON \text{ and } (K1, K'2) OFF \end{cases}$$

### 2.2. Mathematical modeling of the inverter using LC filter

Math equations can be written via node/cycle rules:

(1) $L \frac{di_L}{dt} = v_{AB} - v_S$      (2) $C \frac{dv_S}{dt} = i_L - i_S$

The output voltage from $v_{AB}$ converter can have two values depending on the switching mode; thus, it has the µ control signal.

$v_{AB} = \begin{cases} E \text{ when only } (K1, K'2) \text{ are ON ie } \mu = 1 \\ -E \text{ when only } (K2, K'1) \text{ are ON ie } \mu = -1 \end{cases}$

*t*herefore, resulting in:
(3) $v_{AB} = \mu E$

This switching model is a system with a variable structure: $v_{AB}$ is not a continuous variable; it can have two discontinuous values of E and –E. Consequently, it does not suit an uninterrupted control law design. To solve this problem, the average model has usually been applied (extensively for modeling of stable converters). The proposition (which is verified for the most part in this case) is that the switching period is far shorter, as compared to the system's dynamics. As a result, the formulation is as follows:

(4) $C \dot{x}_1 = x_2 - i_S$
(5) $L \dot{x}_2 = u_E - x_1$

where $x1$ and $x2$ represent the medium values in a period of sampling taken from Vs output voltage of the two ends of C capacitor and iL denotes the current in L inductor. The values for the control variable stand between 1 and -1, indicative of the medium value of the µ control signal, which is formed by the modulated square-shaped pulse width [13-23].

## III. CONTROLLER DESIGN

This is a multi-purpose research: design of three controllers to allow the converter to provide a sinusoidal voltage with a completely fixed frequency and amplitude irrespective of load. The output signal must be a *function of a reference signal* $x1(t) = V\sqrt{2} \sin(\omega t)$
where V= 230V and *f*= 50HZ (ω = 2πf), denoting RMS value and frequency of the reference sinusoidal wave signal, respectively.

### 3.1. Designing controller as sliding mode

In control theory, motion-state control is a nonlinear control method that changes the nonlinear controller method via a control signal which is not continous, which makes the system "move and slide" along a psychograph of the system's natural characteristics. The law belong to the feedback-state is claimed that this controller is not a continuous function of time. Therefore, it can switch from one status structure to another depending status on the position of current in the state-space. As a result, the motion-state control is a variable structure control method. Direct the trajectory towards an adjacent area can considered as Multiple-control structures, which has a different control structure; thus no final trajectory would completely exist within a control structure. So, it will move among the boundaries of control structures. The sliding motion of the system along the boundaries is called a sliding mode and the geometrical locus composed of these boundaries is called the sliding surface [11] [12].

Sliding mode controllers are applied (as an example) for controlling electrical drives which are operated by replacing power converters. A variety of studies have been directed towards this subject.

The purpose of control is always the same: the designed controller must always allow the UPS converter to provide the system with sinusoidal voltage with a completely constant frequency and amplitude irrespective of load. The output voltage must be a function of the reference signal:

$$x_1^*(t) = V\sqrt{2} \sin(\omega t)$$

**Step 1:**

Z1 error is defined as follows:
(6)   $z_1 = C(x_1 - x_1^*)$
Its dynamic is obtained via the following relationships:
(7)   $\dot{z}_1 = C(\dot{x}_1 - \dot{x}_1^*)$
(8)   $\dot{z}_1 = x_2 - i_S - C\dot{x}_1^*$

Given the Lyapunov function as bellow,
(9)   $V_1 = \frac{1}{2} z_1^2$
(10)  $V'_1 = \dot{z}_1 z_1$

And choosing:
$\dot{z}_1 = -k_1 z_1$  →   $z_1(t) = z_1(0)e^{-k_1 t}$

Where k1 is a positive constant value and leads to a Lyapunov function with a negatively defined dynamic; this formulation is made:

(11) $V'_1 = -k_1 z_1^2$

Thus, the asymptotic stability is obtained and Z1 is driven to zero in an ascending manner. In the system (7), x2 is similar to a virtual control input. As a consequence, Z1 can be stabilized at zero if

Then, from (8) and (10), these formulas are obtained

(12) $x_2^* = -k_1 z_1 + i_S + C\dot{x}_1^*$

Where x2* is deemed as a constant function. A new error variable is obtained between virtual drive and its desirable value

(13) $z_2 = x_2 - x_2^*$

From the equations (7), (12) and (13), it can be concluded that

(14) $\dot{z}_1 = -k_1 z_1 + z_2$

**Step 2:**

The derivative of Z2, which is computed with respect to change to time, is given by:

(15) $\dot{z}_2 = \dot{x}_2 - \dot{x}_2^*$
(16) $\dot{z}_2 = \frac{1}{L}(uE - x_1) - \dot{x}_2^*$

Considering a real time control system, the problem regarding the stabilization of the already explained system in equations (14) and (16) can be understood by means of the Lyapunov function below:

(17) $V_2 = \frac{1}{2} z_1^2 + \frac{1}{2} z_2^2$
(18) $V'_2 = \dot{z}_1 z_1 + \dot{z}_2 z_2$
(19) $V'_2 = -k_1 z_1^2 + z_2(z_1 + \dot{z}_2)$
(20) $(z_1 + \dot{z}_2) = -k_2 z_2$
(21) $V'_2 = -k_1 z_1^2 - k_2 z_2^2 < 0$

**Step 3:**

Using (6) Z error is defined as follows:

(22) $z = C(x_1 - x_1^*)$

Its dynamic is obtained via the following relationships:

(23) $\dot{z} = C(\dot{x}_1 - \dot{x}_1^*)$
(24) $\dot{z} = x_2 - i_S - C\dot{x}_1^*$
(25) $\ddot{z} = \frac{1}{L}(uE - x_1) - \frac{di_S}{dt} - C\ddot{x}_1^*$

The sliding surface is defined as follows:

(26) $s(x) = k z + \frac{dz}{dt}$

The dynamics of this sliding function is:

(27) $\dot{s}(x) = k \frac{dz}{dt} + \frac{d^2z}{dt^2}$

Using (24), (25) and (27), we obtain

(28) $\dot{s}(x) = k(x_2 - i_S - C\dot{x}_1^*) + \frac{1}{L}(uE - x_1) - \frac{di_S}{dt} - C\ddot{x}_1^*$

To guarantee the convergence which is belong to the sliding surface in finite time, can define the following Lyapunov function:

$V_1 = \frac{1}{2} s^2 \quad and \quad V'_1 = \dot{s}s$

The derivative of the Lyapunov function is calculated in order to figure out the dynamic while guarantee the derivative of the Lyapunov function is negative. As a result,

(29) $\dot{s} = -\beta\, sgn(s) \rightarrow V'_1 = -\beta|s| < 0$

Using (21) and (28) we can deduce the control law:

(30) $u = x_1 E + \frac{L}{E}(-\beta\, sgn(s) - k(x_2 - i_S - C\dot{x}_1^*)) + LE(\frac{di_S}{dt} + C\ddot{x}_1^*)$

### 3.2. Designing the fuzzy logic controller

*Fuzzy* logic (FL) is a direct, intuitive control system methodology with no *dependence* on mathematical modeling. It receives specific data from different sensors and converts them into functions with fuzzy identities via fuzzification process, during which the data are processed and sorted drawing on a fuzzy (If-Then) rule set in an inference motor. To arrive at a definite conclusion, the fuzzy output is assigned a crisp value through defuzzification process. Fuzzy controller is implemented via fuzzy logic toolbox for the inverter system. This toolbox allows the production of input membership functions, fuzzy control law and output membership functions [14].

To implement the fuzzy controller, the system must have two inputs and one output (as shown in Fig.2) as well as a triangular membership function (Fig. 3).

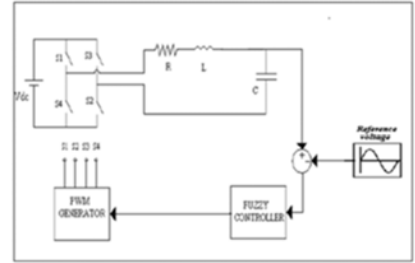

Fig.2. Schematic of a single-phase inverter with a fuzzy controller

Fig.3. Fuzzy membership functions for the inputs and output

where PB stands for positive big, PM for positive medium, PS for positive small, Z for zero and NB for negative big,

[1] The *I*nput is shown as error (e) and change in error as Δe

The membership function A7*7 has 49 member functions; the rules are depicted in table 1:

**Table 1:** fuzzy membership function rules

| E\CE | NB | NM | NS | Z | PS | PM | PB |
|---|---|---|---|---|---|---|---|
| NB | NB | NB | NB | NB | NM | NS | Z |
| NM | NB | NB | NB | NM | NS | Z | PS |
| NS | NB | NB | NM | NS | Z | PS | PM |
| Z | NB | NM | NS | Z | PS | PM | PB |
| PS | NM | NS | Z | PS | PM | PB | PB |
| PM | NS | Z | PS | PM | PB | PB | PB |
| PB | Z | PS | PM | PB | PB | PB | PB |

## IV. SIMULATION RESULT

To validate the proposed method and evaluate the performance, it is employed on either SI (MKS) or CGS as primary units.

In this section, simulation of MATLAB software is implemented for the proposed systems and studies on non-linear back-stepping, sliding and fuzzy mode controllers are analytically compared with regard to the impact of harmonic distortion on voltage and load current.

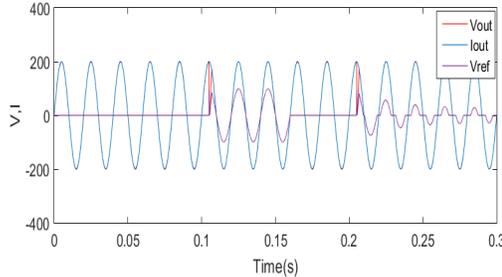

**Fig.5.** Inverter simulation by using the sliding mode controller

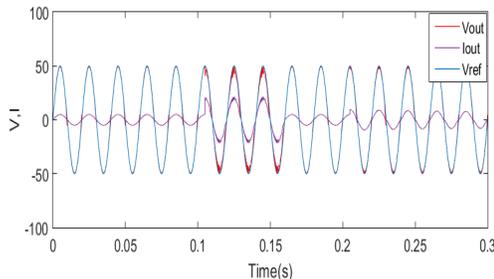

**Fig.6.** Inverter simulation by using the fuzzy mode controller

## V. CONCLUSION

This paper examined a stable DC-AC converter in EPS. Two aims were pursued:

*1.* Designing FLC using three different methods. The control law, which stabilizes the system, provided a perfect tracking of the output voltage (its reaching a desirable value) of a variable belong to the sinusoidal reference and ensured good adjustments in the demonstrate the robustness variety of loads.

*2.* Generating an analytical comparison of the performance and yield of the two controllers:
Sliding and fuzzy modes.

According to the simulation results, every two designed controllers provide a good *robust tracking* of the output voltage. The sliding mode controller is robust but hardly gives any well-defined orders. Fuzzy controller has a better response time as compared to the other but it is deficient in that when there is a change in the loads or in the reference source, the existent current in the inductor is not controllable. However, it is to note that the synthesis of a good-yielding Sliding mode controller is much more complicated than a fuzzy one. In this research, this synthesis was done at different stages with regard to the given control system. The ki constants were selected in an approximate way after many tests were conducted.